# Human Y-chromosome gene classification using Fractal Dimension & Shannon Entropy

Todd Holden* and JianMin Ye
Department of Physics, Queensborough Community College of the City University of New York, Bayside, NY 11364

*Corresponding author: Todd Holden, Department of Physics, Queensborough Community College of the City University of New York, Bayside, NY 11364, e-mail: tholden@qcc.cuny.edu

*Abstract*—**All genes on the human Y-chromosome were studied using fractal dimension and Shannon entropy. Clear outlier clusters were identified. Among these were 6 sequences that have since been withdrawn as CDSs and 1 additional sequence that is not in the current assembly. A methodology for ranking the sequences based on deviation from average values of FD and SE was developed. The group of sequences scored among the 10% largest deviations had abnormally high likelihood to be from centromeric or pseudoautosomal regions and low likelihood to be from X-chromosome transposed regions. lncRNA sequences were also enriched among the outliers. In addition, the number of expressed genes previously identified for evolutionary study tended to not have large deviations from the average.**

## I.  INTRODUCTION

The sequencing of the human genome has led to a wealth of genetic information that has led to many breakthroughs, although most of the genes discovered have unknown or poorly understood functions.  At the same time, the recent explosion in sequencing throughput has enabled studies of populations of a species, normal as opposed to cancerous cells, or even one human DNA sequence to another.  Such studies will require the development of better tools to annotate and classify DNA sequences rapidly [1].

One set of particularly interesting DNA from an evolutionary standpoint is the Y-chromosome. It starts with a relatively short sequence it shares with the X-chromosome. The remaining 95% is best described as a mosaic of complex and interrelated sequences [2]. This includes large duplicative sections that are related to the X-chromosome to some extent (in some cases distantly in others closely), which can be categorized as X-transposed, X-degenerate, and ampliconic. In the ampliconic regions, a number of repetitive palindromic sections appear to facilitate gene transfer between the palindromic arms.  This leads to the possible formation of new genes, which have their origin in recombination between non-critical regions of the genome [3]. On the other hand, the Y-chromosome also houses some powerful genes, such as the 897 bp SRY gene, which alone can control whether an individual is phenotypically male or female [4].

These unique aspects of the Y-chromosome are promising for the study of novel and evolving DNA sequences. Studying the evolution of the Y-chromosome also requires unique and varied tools due to the fact that evolutionary pressure takes a different form due to its unique lack of a partner [5].  The large swaths of the Y-chromosome apparently devoted genomic rearrangement can play an important role in genome evolution [6] and disease [7]. Additionally, the Y-chromosome is uniquely affected by X-to-Y gene conversion, which has been proposed as a way to increase genetic diversity and to resist degradation during evolution [8].

Much of the analysis accomplished for evolution of the Y-chromosome have relied on new and revolutionary models of DNA mutation such as the biological clocks developed over the last several decades [9].  Many comparative models and tools based on aligning base pairs within the genetic code and then using simple empirical rules about single base pair mutation rate have aided in this great effort [10].  For example, the popular BLAST algorithm begins by finding close matches to DNA code segments nine base pairs at a time, and then through an iterative process arranges these segments into finds reasonable candidate alignments, which can then be assigned a comparative score [11].  Only after the alignments have been made can the model be introduced identify relationships between genes.  In phylogenetic analysis, it has been noted that the method of alignment has a bigger effect on derived phylogenetic topology than the method of phylogenetic reconstruction used [12].  In addition, although various algorithms have been employed to narrow the relevant DNA sections to align, these algorithms remain time consuming when one considers whole genome studies [13-15].

With the unique aspects of the Y-chromosome in mind, we have sought to develop a new way to quickly review and curate the genome according to its bioinformatics properties.  Some such metrics have been developed to look at overall characterization of types of DNA (i.e. promoter, interon, exon, etc.) or species. These metrics have included fractal dimension [16,17], Hurst number [18], percentage of $C_pG$ di-nucleotide [19], and percentage all di-nucleotide pairings [20,21].  To better identify groups of genes and genetic outliers, we present in this article a method of finding classes of genes according to their information content as measured by fractal dimension (FD) and Shannon entropy (SE).

## II. METHODS

### A. Genetic Sequence

The complete y-chromosome genome was downloaded using the UCSC Genome Browse. Genome Reference Consortium Human Build 37 patch release 2 (GRCh37.p2) was used. Annotations from The NCBI Eukaryotic Genome Annotation Pipeline were used to identify genes within the sequence. Genes shorter than 75 base pairs were excluded from analysis, leaving a total of 426 genes for analysis.

### B. Higuchi Fractal Method

The ATCG sequence was converted to a numerical sequence, $I$, by assigning the atomic number, the total number of protons, in each nucleotide: A(70), T(66), C(58), G(78). The assigned number is proportional to the nucleotide mass (ignoring isotopes). The A-T and C-G pairs in double stranded DNA have the same value of 136. Among the various fractal dimension methods, the Higuchi fractal method is well suited for studying signal fluctuation and has been applied to nucleotide sequences [22]. The numerical sequence, $I$, is used to generate a difference series ($I(j)-I(i)$) for different lags. The non-normalized apparent length of the series curve is simply $L(k) = \Sigma |I(j)-I(i)|$ for all pairs where ($j-i$) equals $k$. The number of terms in a $k$-series varies and normalization must be used [11]. If $I(i)$ is a fractal function, then a plot of $\log(L(k))$ versus $\log(1/k)$ will be a straight line with the slope equal to the fractal dimension. Higuchi incorporated a calibration division step (division by $k$) such that the maximum theoretical value is calibrated to the topological value of 2 and a minimum value of 1.

### C. Shannon Entropy Calculation

The Shannon entropy of a sequence can be used to monitor the level of functional constraints acting on the gene [23]. A sequence with a relatively low nucleotide variety would have a low Shannon entropy (more constraint) in terms of the set of 16 possible di-nucleotide pairs. A sequence's entropy can be computed as the sum of $\Sigma p_i \log(p_i)$ over all states $i$ and the probability $p_i$ can be obtained from the empirical histogram of the 16 di-nucleotide-pairs. The maximum di-nucleotide entropy is 4 binary bits per pair for 16 possibilities ($2^4$). We use the di-nucleotide Shannon entropy as it preserves some of the sequence order information.

## III. RESULTS AND DISCUSSION

Figure 1 shows a plot of the fractal dimension (FD) vs Shannon information entropy (SE) for all 426 Y-chromosome genes analyzed. Several outliers are marked based on their apparent function. A group of six outlying sequences have been withdrawn as gene candidates. An additional sequence is not in the current assembly. In addition outlying lncRNA (long non gene coding RNA) and pseudogenes are marked.

To make a more quantitative analysis of the data, the data was analyzed for trends and patterns. Figure 2 shows fractal

The project was partially supported by several CUNY grants.

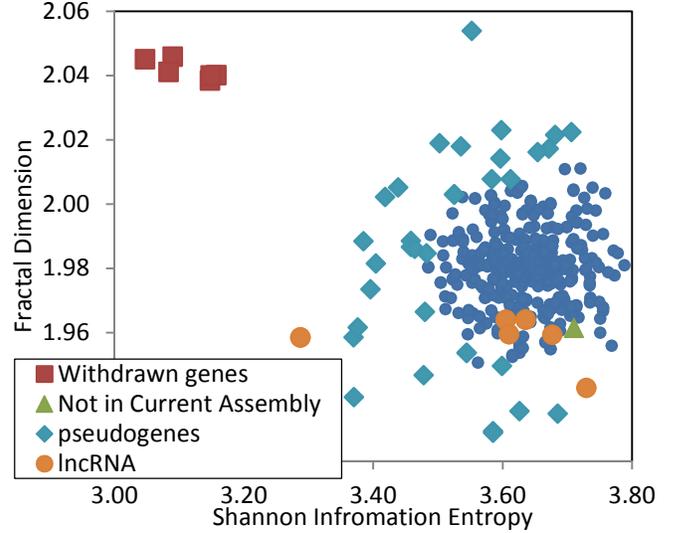

**Figure 1**. Fractal dimension vs Shannon information entropy for all 426 Y-chromosome exons analyzed. Identified outliers are described according to the legend.

dimension and Shannon entropy as functions of sequence length. While the fractal dimension shows a clear reduction in variability, approximately inverse to the square root sequence length, the Shannon entropy shows no clear dependence on sequence, especially if the 6 largest outliers (which have been withdrawn as gene candidates) are neglected. To account for these trends, a weighted distance from average (WDFA) was calculated using the following formula:

$$WDFA = dist_{FD} + dist_{SE}$$
$$= \sqrt{length}(FD - FD_{ave})^2 + (SE - SE_{ave})^2$$

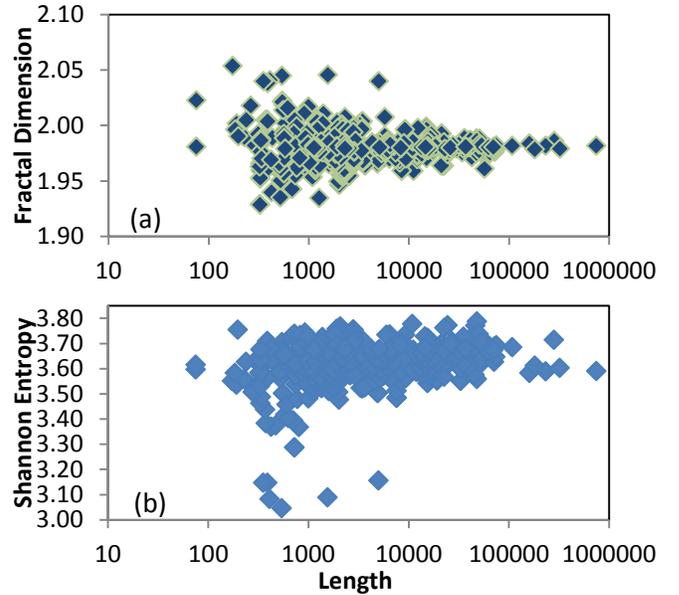

**Figure 2**. (a) Fractal dimension and (b) Shannon information entropy versus length for all 426 Y-chromosome genes analyzed.

where $dist_{FD}$ is the contribution due to deviation from the average FD, $dist_{SE}$ is the contribution due to deviation from the average SE, *length* is the length of the sequence, FD and SE are the fractal dimension and Shannon entropy of the sequence, repectively, and $FD_{ave}$ and $SE_{ave}$ are the average FD and SE for all Y-chromosome gene sequences, respectively.

Figure 3 shows a plot of $dist_{FD}$ vs. $dist_{SE}$. Comparison with Fig. 1 shows that we have improved sensitivity to longer sequences at the cost of reducing the grouping of some similar sequences. The outliers marked in Figs. 1 and 3 are the 43 sequences (10% of the total) with the highest WDFA.

Of these 43 sequences, the National Center for Biotechnology Information (NCBI) gene database lists 6 as withdrawn, 1 as not in the current assembly, 9 as lncRNA (long non-gene coding RNA), and 27 as pseudogenes. None of these are listed in the NCBI database as known gene-coding sequences. This "top 10%" group includes 14 of 38 (37%) of genes near the centromere genes (9.5-14 MB), 2 of 11 (22%) of genes in the pseudoautosomal region, and 9 of 44 (20%) of long non-protein coding RNA (lncRNA) genes. On the other hand, only one sequence from the x-transposed region (3.8%) is present in the group. For completeness, details of these statistics are listed in Table I. (shown at the end after the references)

The inclusion of nine lncRNA sequences among the highest WDFA genes is intriguing. lncRNAs are becoming known for the role they play in tumorgenesis and are a promising area of research for both their role as cancer promoters and supressors [24, 25]. They also have been implicated in developmental processes and stress adaptation [26]. lncRNAs in introns have been shown to have lower FD compared to protein coding exons within the same gene [27].

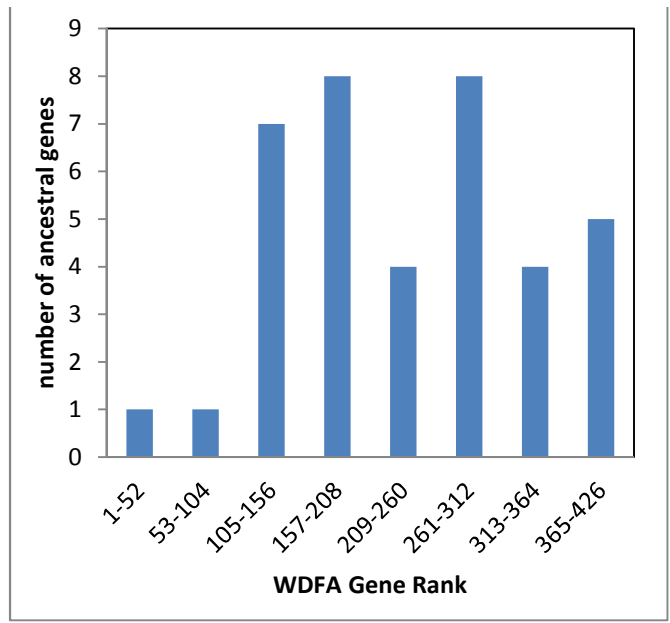

**Figure 4**. Histogram of ancestral genes versus WDFA gene ranking. The vertical axis shows the number of the 38 ancestral Y-chromosome genes existent since before human and old world monkey lineage diverged about 25 million years ago from Ref. [29]. The horizontal axis shows ranges of 426 Y-chromosome genes as ranked by WDFA.

The predominance of near-centromere genes may be related to the mechanism by which they form. It was shown that 30% or more of the centromeric transition region originated from euchromatic gene-containing segments being transposed toward pericentromeric regions [28]. It has been estimated that this type of process has occurred 6-7 times per million years in primate evolution. Having formed recently and via a reduplicative process is likely to increase gene similarity.

To further assess the usefulness of using WDFA to target dysfunctional, evolving, or non-critical genes, we made an analysis of the a previously identified set of 38 ancestral Y-chromosome genes existent since before human and old world monkey lineage diverged about 25 million years ago [29]. Of those genes, very few had high WDFA, with only one in the top decile and one in the second decile. A histogram of these genes vs. WDFA is given in Figure 4. The implication is that a high WDFA score indicates genes that have little or no selection pressure to keep them from evolving, and thus most likely to contribute to novel evolutionary changes.

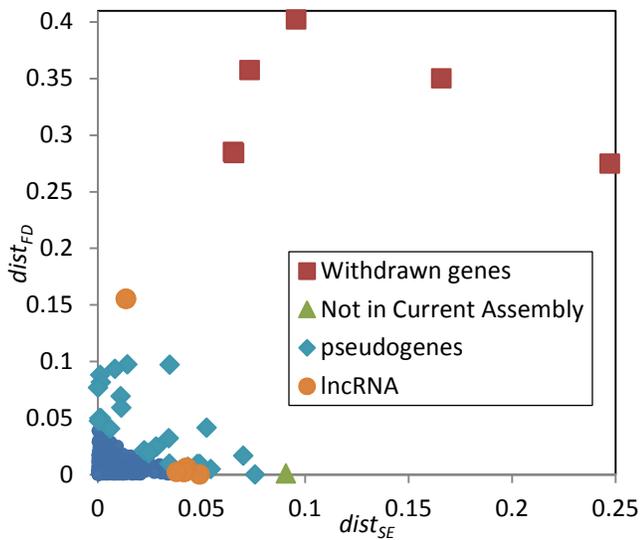

**Figure 3**. Fractal dimension vs Shannon information entropy for all 426 Y-chromosome exons analyzed. Identified outliers are described according to the legend.

IV.  SUMMARY AND CONCLUSION

Our novel analysis of human Y-chromosome genes has succeeded in identifying and partially categorizing genes of interest. A measure (WDFA) was developed to quantify dissimilarity of genes from typical norms. Sensitivity to falsely identified genes, centromeric genes, lncRNA, and pseudoautosomal genes has been demonstrated. Comparison with previously identified highly conserved genes showed that these genes have a clear tendency toward lower WDFA. Further detailed analysis of these results will likely be illuminated as more information becomes available about Y-chromosomal genes. Future studies on other chromosomes will likely show somewhat differing trends of outlying genes due to the unique evolutionary history of the Y-chromosome.

ACKNOWLEDGEMENTS

J.Y. thanks the NSF-REU program for past student support. The authors thank the research groups cited in this paper for posting their data and software in the public domain.

REFERENCES

[1] Alexander RP, Fang G, Rozowsky J, Snyder M, and Gerstein MB (2010) Annotating non-coding regions of the genome. Nature Reviews Genetics 11: 559-571.

[2] Skaletsky H, Kuroda-Kawaguchi T, Minx PJ, Cordum HS, Hillier L, et al. (2003) The male-specific region of the human Y chromosome is a mosaic of discrete sequence classes. Nature 423: 825-837.

[3] Lupski JR, and Stankiewicz P (2005) Genomic disorders: molecular mechanisms for rearrangements and conveyed phenotypes. PLoS Genet. 1: e49.

[4] Chen YS, Racca JD, Phillips NB, and Weiss MA (2013). Inherited human sex reversal due to impaired nucleocytoplasmic trafficking of SRY defines a male transcriptional threshold. Proceedings of the National Academy of Sciences 110(38): E3567-E3576.

[5] Bellott DW and Page DC (2009) Reconstructing the Evolution of Vertebrate Sex Chromosomes. Cold Spring Harb Symp Quant Biol 74: 345-353.

[6] Cheng Z, Ventura M, She X, et al. (2005) A genomewide comparison of recent chimpanzee and human segmental duplications. Nature 437: 88–93.

[7] Bailey JA and Eichler EE. (2006) Primate segmental duplications: crucibles of evolution, diversity and disease. Nat Rev Genet. 7: 552–564.

[8] Trombetta B, Cruciani F, Underhill PA, Sellitto D, and Scozzari R (2010) Mol. Biol. Evol. 27(3): 714–725.

[9] Arbogast BS, Edwards SV, Wakeley J, Beerli P, and Slowinski JB (2002) Estimating divergence times from molecular data on phylogenetic and population genetic timescales. Annu. Rev. Ecol. Syst. 33: 707–40.

[10] Liò P and Goldman N (1998) Models of molecular evolution and phylogeny. Genome Res. 8: 1233-1244.

[11] Altschul SF, Madden TL, Schäffer AA, Zhang J, Zhang Z, Miller W, and Lipman DJ (1997) Gapped BLAST and PSI-BLAST: a new generation of protein database search programs. Nucleic Acids Res. 13: 645–656.

[12] Phillips A, Janies D, and Wheeler W (2000) Multiple sequence alignment in phylogenetic analysis. Mol. Phylogenet. Evol. 16: 317-330.

[13] Altschul SF, Boguski MS, Gish W, and Wootton J (1994) Issues in searching molecular sequence databases. Nature Gen. 6: 119-129.

[14] Mitchison GJ (1999) A Probabilistic Treatment of Phylogeny and Sequence Alignment. J. Mol. Evol. 49: 11–22.

[15] Ogden TH and Rosenberg MS (2006) Multiple Sequence Alignment Accuracy and Phylogenetic Inference. Syst. Biol. 55: 314–328.

[16] Voss RF (1992) Evolution of long-range fractal correlationss and 1/f noise in DNA base sequences. Phys. Rev. Lett. 68: 3805-3808.

[17] Yu ZG, Vo A, Gong ZM, and Long SC (2002) Fractals in DNA sequence analysis. Chinese Physics 11: 1313-1318.

[18] Liu HD, Liu ZH, Sun X (2007) Studies of Hurst Index for Different Regions of Genes. ICBBE 2007: 238-240.

[19] Saxonov S, Berg P, and Brutlag DL (2006) A genome-wide analysis of CpG dinucleotides in the human genome distinguishes two distinct classes of promoters. Proc. Natl. Acad. Sci. USA 103: 1412-1417.

[20] Gentles AJ and Karlin S (2001) Genome-scale compositional comparisonsin eukaryotes. Genome Res. 11: 540-546.

[21] Goldman N (1993) Nucleotide, dinucleotide and trinucleotide frequencies explain patterns observed in chaos game representations of DNA sequences. Nucl. Acids. Res. 21: 2478-2491.

[22] Higuchi T (1998) Approach to an irregular time series on the basis of fractal theory. Physica D 31: 277-283.

[23] Parkhomchuk DV (2006) Di-nucleotide Entropy as a Measure of Genomic Sequence Functionality arXiv:q-bio/0611059.

[24] Gibb EA, Brown CJ, and Lam WL (2011) The functional role of long non-coding RNA in human carcinomas. Molecular Cancer 10(38): 17 pages.

[25] Qi P and Du X (2013) The long non-coding RNAs, a new cancer diagnostic and therapeutic gold mine. Modern Pathology 26, 155–165.

[26] Amor BB, Wirth S, Merchan F, Laporte P, d'Aubenton-Carafa Y, et al. (2009) Novel long non-protein coding RNAs involved in Arabidopsis differentiation and stress responses. Genome Res. 19: 57-69.

[27] Holden T, Nguyen A, Lin E, Cheung E, Dehipawala S, et al. (2013) Exploratory Bioinformatics Study of lncRNAs in Alzheimer's Disease mRNA Sequences with Application to Drug Development. Computational and Mathematical Methods in Medicine 13: Article ID 579136, 8 pages.

[28] She X, Horvath JE, Jiang Z, Liu G, Furey TS, Christ L, et al. (2004) The structure and evolution of centromeric transition regions within the human genome. Nature 430(7002): 857-864.

[29] Hughes, JF, Skaletsky H, Brown LG, Pyntikova T, Graves T, Fulton RS, et al. (2012) Strict evolutionary conservation followed rapid gene loss on human and rhesus Y chromosomes. Nature 483(7387): 82-86.

**Table 1**. Details of genes in the top decile of weighted distance from average (WDFA). The first 4 columns are data from the NCBI Gene database. The last two columns are the Shannon information entropy, fractal dimension, and WDFA calculated for this study.

| Gene Symbol | Description | RefSeq status | Gene Type | Length | SE | FD | WDFA (x1000) |
|---|---|---|---|---|---|---|---|
| LOC100505489 | hypothetical LOC100505489 | withdrawn 8/3/2011 | | 4982 | 3.16 | 2.04 | 522 |
| LOC100505542 | hypothetical LOC100505542 | withdrawn 8/3/2011 | | 1544 | 3.09 | 2.05 | 516 |
| LOC100505575 | hypothetical LOC100505575 | withdrawn 8/3/2011 | | 540 | 3.05 | 2.05 | 498 |
| LOC100507667 | hypothetical LOC100507667 | withdrawn 8/3/2011 | | 408 | 3.08 | 2.04 | 431 |
| LOC100506623 | hypothetical LOC100506623 | withdrawn 8/3/2011 | | 389 | 3.15 | 2.04 | 350 |
| LOC100505513 | hypothetical LOC100505513 | withdrawn 8/3/2011 | | 353 | 3.15 | 2.04 | 350 |
| LOC100287470 | non-protein coding RNA 265C | inferred | lncRNA | 724 | 3.29 | 1.96 | 169 |
| LOC100287296 | non-protein coding RNA 265B | inferred | lncRNA | 724 | 3.29 | 1.96 | 169 |
| MTND6P1 | MTND6 pseudogene 1 | inferred | pseudo | 424 | 3.37 | 1.94 | 132 |
| ZNF884P | zinc finger protein 884, pseudogene | inferred | pseudo | 803 | 3.37 | 1.96 | 111 |
| MTND2P3 | MTND2 pseudogene 3 | inferred | pseudo | 470 | 3.38 | 1.96 | 101 |
| LOC100132421 | double homeobox 4 like 18 | inferred, dux4l18 | pseudo | 2012 | 3.48 | 1.95 | 94 |
| LOC100132230 | family with sequence similarity 201, member C | not in current assembly | | 56706 | 3.71 | 1.96 | 92 |
| CYorf1 | OFD1 pseudogene 18 | inferred | pseudo | 375 | 3.38 | 1.99 | 89 |
| MED13P1 | mediator complex subunit 13 pseudogene 1 | inferred | pseudo | 173 | 3.55 | 2.05 | 87 |
| ZNF736P8Y | zinc finger protein 736 pseudogene 8, Y-linked | inferred | pseudo | 702 | 3.40 | 1.97 | 83 |
| C2orf27AP1 | chromosome 2 open reading frame 27A pseudogene 1 | inferred | pseudo | 586 | 3.42 | 2.00 | 80 |
| ZNF885P | zinc finger protein 885, pseudogene | inferred | pseudo | 629 | 3.40 | 1.98 | 77 |
| LOC140028 | glyceraldehyde 3 phosphate dehydrogenase pseudogene 19 | inferred | pseudo | 1267 | 3.68 | 1.93 | 76 |
| MTND1P1 | MTND1 pseudogene 1 | inferred | pseudo | 368 | 3.44 | 2.01 | 70 |
| LOC100506481 | hypothetical LOC100506481 | model | pseudo | 557 | 3.50 | 2.02 | 66 |
| ARSFP1 | arylsulfatase F pseudogene 1 | inferred | pseudo | 5736 | 3.61 | 2.01 | 59 |
| TCEB1P17 | transcription elongation factor B (SIII), peptide 1 pseudogene 17 | validated | pseudo | 326 | 3.59 | 1.93 | 57 |
| USP9YP1() | USP9Y pseudogene 1 | validated | pseudo | 3339 | 3.52 | 2.00 | 53 |

| Gene Symbol | Description | RefSeq status | Gene Type | Length | SE | FD | WDFA (x1000) |
|---|---|---|---|---|---|---|---|
| USP9YP2() | ubiquitin specific peptidase 9, Y-linked pseudogene 2 | validated | pseudo | 3339 | 3.52 | 2.00 | 52 |
| LOC359998 | CHRFAM7A pseudogene 2 | validated | pseudo | 606 | 3.46 | 1.99 | 51 |
| LOC359999 | CHRFAM7A pseudogene 1 | validated | pseudo | 606 | 3.46 | 1.99 | 51 |
| LOC643001 | double homeobox 4 like 17 | inferred, dux4l18 | pseudo | 1967 | 3.60 | 1.95 | 50 |
| LOC100419952 | hypothetical LOC100419952 | inferred | pseudo | 520 | 3.63 | 1.94 | 50 |
| TTTY13 | testis-specific transcript, Y-linked 13 (non-protein coding) | provisional | lncRNA | 11067 | 3.68 | 1.96 | 49 |
| LOC100101117 | testis-specific transcript, Y-linked 2B (non-protein coding) | validated | lncRNA | 22201 | 3.60 | 1.96 | 49 |
| TTTY2 | testis-specific transcript, Y-linked 2 (non-protein coding) | validated | lncRNA | 22191 | 3.60 | 1.96 | 48 |
| TCEB1P14 | transcription elongation factor B (SIII), peptide 1 pseudogene 14 | validated | pseudo | 331 | 3.46 | 1.99 | 48 |
| TTTY7B | testis-specific transcript, Y-linked 7B (non-protein coding) | validated | lncRNA | 8439 | 3.61 | 1.96 | 47 |
| TTTY7 | testis-specific transcript, Y-linked 7 (non-protein coding) | validated | lncRNA | 8439 | 3.61 | 1.96 | 47 |
| LOC360022 | PC4 and SFRS1 interacting protein 1 pseudogene 2 | inferred | pseudo | 773 | 3.48 | 1.97 | 46 |
| CDY7P() | chromodomain protein, Y-linked 7 pseudogene | validated | pseudo | 2297 | 3.58 | 2.01 | 44 |
| LOC100101116 | testis-specific transcript, Y-linked 1B (non-protein coding) | validated | lncRNA | 21164 | 3.64 | 1.96 | 44 |
| TCEB1P9 | transcription elongation factor B (SIII), peptide 1 pseudogene 9 | inferred | pseudo | 326 | 3.59 | 1.93 | 58 |
| TTTY1 | testis-specific transcript, Y-linked 1 (non-protein coding) | provisional | lncRNA | 21134 | 3.64 | 1.96 | 44 |
| MRP63P10 | mitochondrial ribosomal protein 63 pseudogene 10 | inferred | pseudo | 262 | 3.54 | 2.02 | 44 |
| ZNF736P12Y | zinc finger protein 736 pseudogene 12, Y-linked | inferred | pseudo | 1092 | 3.54 | 1.95 | 43 |
| ZNF736P11Y | zinc finger protein 736 pseudogene 11, Y-linked | inferred | pseudo | 1092 | 3.54 | 1.95 | 43 |